\documentclass[aps,onecolumn,showpacs,superscriptaddress,preprintnumbers,amsmath,amssymb,pra,10pt]{revtex4-2}
\usepackage{graphicx}
\usepackage{dcolumn}
\usepackage{appendix}
\usepackage{bm}
\usepackage{bold-extra}
\usepackage[T1]{fontenc}
\usepackage[usenames,dvipsnames]{xcolor}
\usepackage{subfigure}
\usepackage{bbm}
\usepackage{enumerate}

\definecolor{Cerulean}{rgb}{0.,0.59,0.835}
\definecolor{RubineRed}{rgb}{0.61,0.07,0.12}
\definecolor{myblue}{rgb}{0.2,0.2,0.8}
\usepackage[colorlinks=true,citecolor=myblue,linkcolor=RubineRed,urlcolor=Cerulean]{hyperref}

\usepackage{titlesec}
\usepackage{array}
\usepackage{dsfont}
\usepackage{physics}
\usepackage{mathtools}
\usepackage{upgreek}
\newcommand{\pa}[1]{\left( #1\right)}
\newcommand{\co}[1]{\left[ #1\right]}
\newcommand{\stx}[1]{_\text{#1}}

\begin{document}

\title{Quantum Kramers-Henneberger Transformation}
\author{Javier Arg\"uello-Luengo}
\email{javier.arguello.luengo@upc.edu}
 \thanks{Corresponding author.}
\affiliation{Departament de Física, Universitat Polit\`ecnica de Catalunya, Campus Nord B4-B5, 08034 Barcelona, Spain}
\affiliation{Institute of Photonic Sciences, The Barcelona Institute of Science and Technology, 0886 Castelldefels, Spain}

\author{Javier Rivera-Dean}
\email{javier.rivera@icfo.eu}
\affiliation{Institute of Photonic Sciences, The Barcelona Institute of Science and Technology, 0886 Castelldefels, Spain}

\author{Philipp~Stammer}
\email{philipp.stammer@icfo.eu}
\affiliation{Institute of Photonic Sciences, The Barcelona Institute of Science and Technology, 0886 Castelldefels, Spain}
\affiliation{Atominstitut, Technische Universität Wien, Stadionallee 2, 1020 Vienna, Austria}

\author{Marcelo~F.~Ciappina}
\email{marcelo.ciappina@gtiit.edu.cn}
\affiliation{Department of Physics, Guangdong Technion - Israel Institute of Technology, 241 Daxue Road, Shantou, Guangdong, 515063, China}
\affiliation{Technion - Israel Institute of Technology, Haifa, 32000, Israel}
\affiliation{Guangdong Provincial Key Laboratory of Materials and Technologies for Energy Conversion, Guangdong Technion - Israel Institute of Technology, 241 Daxue Road, Shantou, Guangdong, 515063, China}

\author{Maciej Lewenstein}
\email{maciej.lewenstein@icfo.eu}
\affiliation{Institute of Photonic Sciences, The Barcelona Institute of Science and Technology, 0886 Castelldefels, Spain}
\affiliation{ICREA, Pg. Lluis Companys 23, ES-08010 Barcelona, Spain}

\begin{abstract}
Convenient unitary mappings can distill complex quantum dynamics into an intuitive physical picture that facilitates experimental realizations. The classical Kramers-Henneberger transformation connects the dynamics of a quantum particle of mass $m$ located in a trap at position $\alpha(t)$ with the dynamics of a charge $e$ moving in an electric field $e{\cal{E}}(t)=-m\ddot{\alpha}(t)$ within the dipole approximation. In this paper, we extend the Kramers-Henneberger transformation to the quantum electrodynamic and quantum optical realm by explicitly treating the trap location quantum mechanically, thus taking into account the quantum fluctuations of the time-dependent displacement force. Compared to the classical case, we show that quantum electrodynamic corrections appear, and we propose an optomechanical realization for the quantized position of the trap to show that such corrections can manifest in state-of-the-art experiments. These results open the path to novel quantum simulation of quantum electrodynamics and quantum optics of attoscience and ultrafast physics by using ultracold trapped atoms and ions.
\end{abstract}
\maketitle

\allowdisplaybreaks
\section{Introduction}
The Nobel Prize in Physics recognized in 2023 the achievements reached by atto-science (AS)~\cite{Agostini_Nobel_2024,Krausz_Nobel_2024,Huillier_Nobel_2024}, where an ultrafast laser pulse can be used to induce non-linear dynamics in a target atom, molecule or solid material~\cite{Dra23,JKP11}.~In such processes, the motion of electrons plays a prominent role, and their position is traditionally treated quantum-mechanically to capture fundamental aspects of their dynamics, including optical tunneling~\cite{keldysh_ionization_1965}, phase coherence~\cite{Wituschek2020}, and multi-electron ionization processes~\cite{Krausz2009,MF15,MF16,figueirademorissonfaria_it_2020}.

Regarding the electromagnetic (EM) field, the physics of intense laser-matter interactions was initially based on multiphoton processes, studied via high-order perturbation theory where both matter and light were treated quantum mechanically. With the advent of chirped pulse amplified lasers~\cite{Mourou2019,Strickland2019} producing ultra-intense and ultra-short coherent laser pulses, the need for a quantum electrodynamics (QED) description of the electromagnetic fields lost relevance~\cite{JKP11,ABC19,ashida2021cavity}.~Contemporary AS and, more generally, ultrafast laser physics~\cite{Agostini_Nobel_2024,Krausz_Nobel_2024,Huillier_Nobel_2024}, commonly use the classical description of EM fields while keeping a full quantum description of the material system~\cite{Bra09}. This is sufficient to understand and control~\cite{lewenstein-principles-2008}, the physical principles behind high-harmonic generation (HHG)~\cite{corkum_plasma_1993,krause_high-order_1992,kulander_dynamics_1993,lewenstein-theory-1994}, electron-impact ionization (EII), above-threshold ionization~\cite{Krausz2009}, or multielectron ionization~\cite{ABC19} in atoms, molecules~~\cite{vampa_linking_2015,vampa_semiclassical_2015}, and `simple' solids~\cite{vampa_linking_2015,vampa_semiclassical_2015,ghimire_high-harmonic_2019} in correspondence with experimental observations~\cite{vampa_observation_2018,ghimire_generation_2012,ghimire_observation_2011}. 
This classical approach is also followed in current attempts to characterize quantum materials with attosecond resolution, such as high-$T_c$ superconductors~\cite{ABB22}, Mott insulators or charge-density waves~\cite{TZV24,ZTV24}. 

However, this classical description of the field can be limiting in emission processes~\cite{RS09,GLW10,PGH11,AVS17}, particularly regarding the coherence and entanglement used in attosecond metrology~\cite{BBG20,Vra21,KMD22}, electron-electron correlations in two-electron ionization~\cite{MML22}, or the entanglement between an emitted electron and the parent ion that is probed through photoelectron state characterization~\cite{BLF22,LFD22,LLW23,ALK20}.
To fully describe such phenomena, a QED and quantum optics (QO) treatment of AS has been recently established~\cite{BT19,CDF24,SRM23}, extending to a quantum-optical description of harmonic generation in solids and molecules~\cite{Gorlach2020,rivera-dean_Book_2024,rivera-dean_quantum_2023}, and revealing the coherence, entanglement, and squeezing properties of the emitted harmonic field~\cite{stammer2024absence,Stammer_PRL_2024,Stammer_NatPhys_2024,Stammer_RobustCAT_2024}. 
This framework has further enabled the generation of nonclassical and squeezed light~\cite{YBS24,Gonoskov_PRB_2024,CC22}, with applications in nonlinear optics and photoemission~\cite{Lamprou2020,Lamprou_ArXiv_2023,HMM24,SRE24} that include recent experiments providing a direct evidence of the quantum-optical nature of high-harmonic generation~\cite{TCS24,theidel_observation_2024,Vampa_ArXiv_2024,boroumand_quantum_2025}, thus pushing the development of complementary theoretical models~\cite{Moiseyev_ArXivConditioning_2024,lange_hierarchy_2025,GLM25,WB25,stammer2025theory}.
Other investigations with a quantized light degree of freedom include strong field driven processes induced by non-classical light, such as bright squeezed states~\cite{Kaminer_squeezing,tzur_measuring_2025,rivera2025attosecond, stammer2025weak}, massively entangled quantum states (MQS) of light reconstructed via post-selection methods~\cite{Sta22,rivera-dean_bloch_2023,Stammer_ArXivConditioning_2024,Stammer_EnergyConservation2024}, the generation of entangled states including dipole correlations beyond the low depletion limit~\cite{Stammer_PRL_2024}, and the generation of light-matter entanglement in atoms, molecules, and semiconductors~\cite{RLP22,RSM22,Riv24,CKN24}.

Such studies have shown that the basic building blocks of where and how to find MQS are largely understood. They include: (i) the use of dipole correlations through depletion~\cite{Stammer_PRL_2024}, through superposition states in atoms~\cite{RCS24}, through cavity-induced resonances~\cite{YBS24}, and many other mechanisms in non-trivial materials~\cite{PGR23,lange_electron-correlation-induced_2024,lange_excitonic_2025}; (ii) the use of specific post-selection approaches~\cite{LCP21,RSP21,SRL22}; and (iii) the use of non-classical driving fields~\cite{KK22,GTB22,RCB24,RSC24}, whose engineering is an active challenge that we face in the present work.

For this purpose, quantum simulation of AS using ultra-cold atoms has emerged as a complementary tool to engineer controllable input pulses and gain unprecedented experimental access to the resulting emission. In the classical realism, this approach was stimulated by the work on the Kramers-Henneberger (KH) transformation to the so-called acceleration gauge, in the context of studies of electron stabilization in intense laser fields by M. Gavrila~\cite{Gav67,Gav92,Gav22} and J. Kami\'nski~\cite{GK84}. Following this strategy, an effective dipolar field is induced by the shaking of an optically trapped Bose-Einstein condensate that plays the role of the electron, as proposed in Ref.~\cite{DSS98}.~These ideas have then been further developed theoretically~\cite{AH10, SFS17} and experimentally~\cite{RFS17,SRS18}, as well as extended to the regimes of HHG~\cite{ARS24} and EII in traps shaken or controlled by external time-dependent EM fields~\cite{Maribel}. 

In contrast to this classical control of the trapping coordinate $\alpha(t)$, in this work we investigate quantum simulators of attoscience with cold atoms subjected to a quantized shaking of the trapping position $\hat \alpha (t)$, which requires the derivation of a \textit{quantum Kramers-Henneberger transformation} (QKH). The physics that emerges from this lies within {\it terra incognita} and can open new avenues for quantum simulation and attoscience. For example, how can one significantly advance attoscience further with QED and quantum optics? Can this be simulated using, e.g., ultracold atoms or optomechanical devices? In this work, we address such questions and provide answers towards its realization.

This work is structured as follows. In Section~\ref{secII}, we discuss the classical Kramers-Henneberger transformation, as introduced for electrons in Refs.~\cite{Gav67,Gav92,Gav22,GK84}, or for ultra-cold atoms in Ref.~\cite{DSS98}. 
We generalize this approach to the quantum case in Section~\ref{secIII}, in which the location of the trap is described by a quantum coordinate corresponding to a single quantum mechanical (phononic) oscillator. We generalize this approach in Section~\ref{secIV} to a description where the quantum coordinate is a continuum of oscillator modes. 
In Section~\ref{secNum}, we numerically benchmark the effect of leading-order quantum corrections in the context of HHG, and Section~\ref{secVI} contains a discussion of possible experimental realizations using optomechanical devices. The basic consequences and outlook of the previous results are discussed in Section~\ref{secV}. 
In the Appendix we discuss in more detail approximations and expansions that we use to treat the case of the quantum Kramers-Henneberger transformation.

\section{Classical Kramers-Hennberger transformation}
\label{secII}

The seminal studies by H.A. Kramers~\cite{Kramers} and W.C. Henneberger~\cite{Henneberger} introduced the, so-called, \textit{transformation of the motion of charged particle in the electric field in a dipolar approximation} to a coordinate system moving with the free particle in the same field. This is often termed as a transformation to an acceleration gauge, and is very useful in strong laser physics, as pioneered in the works of M. Gavrila et al.~\cite{Gav67,Gav92,Gav22,GK84}. In fact, here we focus on the inverse KH transformation.

Let us consider a quantum mechanical particle of mass $m$ moving in a trap, described by a potential $V(x)$, (in 1D for simplicity, but generalizations to 2D and 3D are straightforward).~The trap is shaken with a time-dependent displacement described by the classical amplitude $\alpha(t)$, so that its dynamics are governed by the Hamiltonian
\begin{equation}
\label{eq:hamIn}
\hat H= \frac{\hat p^2}{2m} + V(\hat x-\alpha(t)),
\end{equation}
where $\hat p=-i\hbar \partial/\partial x$. The Schrödinger equation for the wave function $\psi_0(x,t)$ is then given by
\begin{equation}
i\hbar \frac{d}{dt} \psi_0(x,t) = \left[\frac{\hat p^2}{2m} + V(\hat x-\alpha(t))\right]\psi_0(x,t).
\end{equation}

We now perform a series of unitary transformations, starting with $\ket{\psi_1(t)}= \hat U_0 \ket{\psi_0(t)}$, where 
\begin{equation}
    \hat U_0 = \exp[i\,\hat p\alpha(t)/\hbar]=\exp[\partial_x\alpha(t)].
\end{equation}

The Schrödinger equation for $\psi_1(x,t)$ reads
\begin{equation}
i\hbar \frac{d}{dt} \psi_1(x,t) = \left[\hat U_0 \hat H \hat U_0^\dagger + i\hbar \dot{\hat U}_0 \hat U_0^\dagger\right]\psi_1(x,t),
\end{equation}
which, after completing the square, becomes 
\begin{equation}
i\hbar \frac{d}{dt} \psi_1(x,t) = \left[ \frac{(\hat p- m \dot \alpha(t) )^2}{2m} -\frac{ m \dot\alpha^2(t)}{2}+ V(\hat x)\right]\psi_1(x,t).
\end{equation}

The term $m \dot \alpha(t)^2/2$ can be removed with the unitary transformation  $\hat U_1=\exp[-\frac{im}{2\hbar}\int_{t_0}^t dt'\,\dot \alpha^2(t')]$.~We then apply a momentum translation, $\hat U_2=\exp[-i \hat x m\dot \alpha(t)/\hbar]$ to obtain the final form of the evolution for $\ket{\psi(t)}= \hat U_2 \hat U_1\ket{\psi_1(t)}$; 
\begin{equation}
i\hbar \frac{d}{dt} \psi(x,t) = \left[\frac{\hat p^2}{2m} + V(\hat x)+m \hat x \ddot\alpha(t) \right]\psi(x,t).
\end{equation}
Clearly, the final Schrödinger equation is equivalent to the one describing the evolution of the electron of charge $e$ in the time-dependent classical electric field ${\cal E}(t)$ in the dipole approximation
\begin{equation}
i\hbar \frac{d}{dt} \psi(x,t) = \left[\frac{\hat p^2}{2m} + V(\hat x)-e \hat x {\cal E}(t) \right]\psi(x,t), 
\end{equation}
upon identification $m \hat x \ddot\alpha(t) = -e \hat x {\cal E}(t)$.

\section{Quantum Kramers-Henneberger Transformation -- single mode case}
\label{secIII}
In this Section we consider again a quantum particle in a trap, but the location $\hat \alpha (t)$ of the trap is quantized: as it occurs for the position operator of a quantum harmonic oscillator. We discuss possible experimental realizations in more detail in Section~\ref{secVI}. Here, we want to generalize the classical KH transformation by treating the position of the trap quantum mechanically. 
For simplicity, we start with the case of a single quantum mode. The Schrödinger equation for the wave function  of the total system $|\phi(t)\rangle$ is then governed by the time-dependent Hamiltonian  
\begin{equation}
\label{eq:quantizedShaking}
i\hbar \frac{d}{dt} \phi(x,t) = \left[ \frac{\hat p^2}{2m} + V(\hat x - \hat \alpha_x(t))+\hbar \omega \hat a ^\dagger \hat a \right]\phi(x,t).
\end{equation} 

Here, $\hat a ^\dagger$ and  $\hat a$ are creation and annihilation operators of the quantum mechanical oscillator, respectively. Note that $|\phi(t)\rangle$ is now a state vector in the total Hilbert space of the particle and the quantum oscillator. 
Similarly as in Ref.~\cite{LCP21}, we introduce an envelope function $f(t)$ to control the finite duration of the shaking  process. The envelope is a smooth function, $0\le f(t)\le 1$, that starts at $t_0$ and ends at $t_f$. These times can be taken to be $\pm \infty$. Also, we may demand that both $f(t)$ and $\partial_t f(t) \equiv f^\prime(t)$ vanish at $t_0$ and $t_f$. 
The quantum trap position then reads 
\begin{equation}
    \hat \alpha_x(t)= \ell f(t)(\hat a +\hat a ^\dagger),
\end{equation}
where $\ell$ is a coupling constant with the dimension of length. 
First, we transform the Schrödinger equation to the interaction picture with the unitary transformation $\hat U\stx{osc}=e^{i\hat H\stx{osc} t/\hbar}$ to account for the free quantum oscillator Hamiltonian $\hat H\stx{osc} = \hbar \omega \hat a^\dagger \hat a$.~The quantum trap location then becomes 
\begin{equation}
    \hat \alpha(t)= \ell f(t)(\hat a e^{-i\omega t} +\hat a ^\dagger e^{i\omega t}).
\end{equation}

The resulting Schrödinger equation for the transformed state $|\psi_0(t)\rangle=\hat U\stx{osc}|\phi(t)\rangle$ reads
\begin{equation}
i\hbar \frac{d}{dt} \psi_0(x,t) = \left[\frac{\hat p^2}{2m} + V(\hat x-\hat \alpha(t)) \right]\psi_0(x,t).
\end{equation}

In the next step we shift the position $\hat x$ to $\hat \alpha(t)$, using the unitary transformation
\begin{equation}
       \hat U_0 = \exp\left(\frac{\partial}{\partial x} \mathcal{T}\int_{t_0}^t dt'\,\dot{\hat\alpha}(t') \right),
\end{equation}
where 
\begin{eqnarray}
    \dot{\hat \alpha}(t)&=& +i\omega \ell f(t)\left(-\hat a e^{-i\omega t}+\hat a^\dagger e^{i\omega t}\right)+\ell f^\prime(t)\left(\hat a e^{-i\omega t}+\hat a^\dagger e^{i\omega t}\right).\nonumber
\end{eqnarray}

Note that we now deal with operators, so that we use time-ordered products. Also, note that when the characteristic scale $T$, on which $f(t)$ varies is moderately slow, $1 \ll \omega T$, the second term in the above expression can be safely neglected.~We use this approximation in the following; although generalizations to the regime $1 \simeq \omega T$ pose only minor technical difficulties.~In general, the transformation above yields
$
    \hat U_0 \hat x \hat U_0^\dag = \hat x + \hat\alpha(t).
$

The resulting Schrödinger equation for $|\psi_1(t)\rangle=\hat U_0 |\psi_0(t)\rangle$ now reads
\begin{equation}
\begin{aligned}
&i\hbar \frac{d}{dt} \psi_1(t)= \left[\frac{(\hat p-m \dot {\hat\alpha}(t) )^2}{2m} -\frac{m \dot {\hat \alpha}(t)^2}{2}+ V(\hat x)\right]\psi_1(x,t).
\end{aligned}
\end{equation}

Unlike the classical scenario, in this case, the term $m \dot {\hat \alpha}(t)^2/{2}$ does not correspond to a trivial phase factor, as it has a non-trivial character in the Hilbert space of the quantum oscillator. Still, it can be eliminated, renormalizing $m \dot {\hat \alpha}(t)$.~However, and importantly, due to its quadratic dependence on the oscillator operators $\hat a$ and $\hat a^\dagger$, it automatically leads to squeezing (stretching) of the initial state of the quantum oscillator. 

To show this, we now introduce $|\psi_2(t)\rangle = \hat U_{1}|\psi_1(t)\rangle$ for the renormalization, which may now lead to squeezing or anti-squeezing,
\begin{equation}
\label{eq:U1}
    \hat U_{1}=\mathcal{T}\exp\left(-\frac{im}{2\hbar}\int_{t_0}^t dt'\, \dot{\hat \alpha}(t')^2\right) \,,
\end{equation}
so that we get
\begin{equation}
i\hbar \frac{d}{dt} \psi_2(x,t) = \left[\frac{(\hat p-m \dot {\hat\alpha}_\text{eff}(t) )^2}{2m} + V(\hat x)\right]\psi_2(x,t).
\end{equation}
for an effective field operator $\hat\alpha_\text{eff}(t)$ that we evaluate in the next paragraph. Finally, we use $|\psi(t)\rangle=\hat U_2 |\psi_2(t)\rangle$, with
\begin{equation}
    \hat U_2=\exp\left(\frac{-im }{\hbar} \hat x \, \mathcal{T}\int_{t_0}^t dt'\,\ddot{\hat \alpha}_\text{eff}(t') \right) \,,
\end{equation}
which brings us to the final result
\begin{equation}
i\hbar \frac{d}{dt} |\psi(x,t)\rangle = \left[\frac{\hat p^2}{2m} + V(\hat x) + m \hat x \ddot{\hat\alpha}_\text{eff}(t)\right]|\psi(x,t)\rangle.
\label{final}
\end{equation}

To calculate the effective field operator $\ddot{\hat\alpha}_\text{eff}(t)$ in Eq.~\eqref{eq:U1}, we observe the following (details can be found in the Appendix~\ref{sec:appendix})
\begin{equation}
\begin{aligned}
&\exp\!\bigg(\dfrac{-i m \dot{\hat \alpha}(t')^2}{2\hbar}\bigg)\ \dot{\hat \alpha}(t)\exp\!\bigg(\dfrac{i m \dot{\hat \alpha}(t')^2}{2\hbar}\bigg)
=   \dot{\hat \alpha}(t) - \frac{i m}{2\hbar}\left[\dot{\hat \alpha}(t')^2, \dot{\hat \alpha}(t)\right]
=   \dot{\hat \alpha}(t) - \frac{i m}{\hbar}F(t',t)\dot{\hat \alpha}(t'),
\end{aligned}
\end{equation}
where $F(t',t) = \left[\dot{\hat \alpha}(t'), \dot{\hat \alpha}(t)\right]$ is a $c$-number.
Note that the characteristic scale $1/\gamma$ on which the interaction duration varies depends entirely on the initial oscillator wave packet. In the regime $\gamma\ll \omega$, we obtain $F(t',t)= 2 i \omega^2\ell^2f(t')f(t)\sin[\omega(t-t')]$, and 
\begin{equation}
\begin{aligned}
\dot{\hat \alpha}_\text{eff}(t)&= \hat U_{1} \dot{\hat \alpha}(t) \hat U_{1}^\dag = \dot{\hat \alpha}(t) - \frac{im}{\hbar}\int_{t_0}^t dt'\, F(t', t)\dot{\hat \alpha}(t') + \rm{...}= \dot{\hat \alpha}(t)\left[1 + O(\epsilon) + O(\epsilon^2) + ...\right],
\end{aligned}
\end{equation}
where $\epsilon =m\omega^2 \ell^2 /(\hbar\omega)$ is `small' in the traditional parameter regimes. The explicit expression is expanded in Appendix~\ref{sec:appendix}.
Interestingly, the smallness of $\epsilon$ is equivalent to,
\begin{equation}
\label{eq:def_epsilon}
    \epsilon=\frac{m \omega^2 \ell^2}{\hbar\omega}=\frac{\ell^2}{2a_\text{zp}^2}\ll 1\,,
\end{equation}
where $a\stx{zp}=\sqrt{\hbar/(2m\omega)}$ is the zero point motion associated with a quantum harmonic oscillator of mass $m$ and frequency $\omega$. Thus, the amplitude of the perturbation $\ell$ needs to satisfy $\ell\ll a\stx{zp}$ for higher-order corrections to be neglected. 
The second derivative entering Eq.~\eqref{final} then expands as,
\begin{eqnarray}
\label{eq:corrections_1freq}
    &&\ddot{\hat\alpha}_\text{eff}(t) \simeq \ddot{\hat\alpha}(t) - \frac{i m}{\hbar}\int_{t_0}^t dt'\, \frac{dF(t', t)}{dt}\dot{\hat \alpha}(t')  \label{alphaR}\\
    &\simeq& \ddot{\hat\alpha}(t) + \frac{2m \ell^2\omega^3}{\hbar} f(t)\int_{t_0}^t dt'\, f(t')\cos[\omega(t-t')]\dot{\hat \alpha}(t') \nonumber \\ \nonumber
    \end{eqnarray}
    
The main result of this Section is contained in Eqs.~\eqref{final} and \eqref{alphaR}. As we can see, the quantum KH transformation leads to the Schrödinger equation for the trapped particle in an effective quantized `electric field' operator, $\hat {\cal E}(t) = -\frac{m}{e}\ddot{\hat \alpha}_\text{eff}(t)$. Interestingly, the corresponding effective field is not simply proportional to $\expval*{\ddot{\hat\alpha}(t)}$, as one could expect from the classical case. Instead, this field includes a quantum electrodynamic correction, as illustrated in Eq.~\eqref{alphaR}. The consequences of the above results are numerically explored in Section~\ref{secNum} and discussed in more detail in Section~\ref{secV}.

\section{Quantum Kramers-Henneberger Transformation -- continuous spectrum of phonons}
\label{secIV}
In this Section we consider again a trapped quantum particle with a quantized location of the trap, but we generalize the previous discussion to the case of a continuous spectrum of harmonic oscillators. This allows us to forget about the phenomenological description of time dependence using the envelope function $f(t)$. Instead, we work from the very beginning with a time-independent Hamiltonian that conserves energy along the quantum dynamics. The temporal shape of the excitations (`laser pulses') is achieved here by combining an appropriate initial wave packet of the oscillator modes with desired temporal and spatial properties, similar to the generalization of a discrete set to a continuum of modes as in Ref.~\cite{SRM23}.

We start with the time-independent Hamiltonian, and with the Schrödinger equation for the state of the total system 
\begin{equation}
\begin{aligned}
    &i\hbar \frac{d}{dt} \phi(x,t)
    = \left[ \frac{\hat p^2}{2m} + V(\hat x-\hat \alpha_x)+\int d\omega\, \hbar \omega \hat a_\omega ^\dagger \hat a_\omega \right]\phi(x,t).
    \end{aligned}
\end{equation} 

Here, $\hat a_\omega^\dagger$ and  $\hat a_\omega$ are creation and annihilation operators of the quantum oscillators of the mode with energy $\hbar \omega$, respectively. We integrate over the index $\omega$, which, in general, is multidimensional and may incorporate polarization or other discrete degrees of freedom. 
Note that, as in Section~\ref{secIII}, $|\phi(t)\rangle$ is a state vector in the total Hilbert space of the quantum oscillators and the trapped particle. The quantum trap location is now given by 
\begin{equation}
    \hat \alpha_x= \int d\omega\, \ell_\omega (\hat a_\omega +\hat a_\omega^\dagger) \,,
\end{equation}
where this time $\ell_\omega$ is a coupling constant density of dimension of length divided by the dimension of $[d\omega]$ (which  depends on the dimension of the oscillator system).
 
As in Sec.~\ref{secIII}, we first transform the Schrödinger equation to the interaction picture with respect to the free quantum oscillator Hamiltonian. The quantum trap location then becomes
\begin{equation}
    \hat \alpha(t)=  \int   d\omega \, \ell_\omega (\hat a_\omega e^{-i\omega t} +\hat a_\omega^\dagger e^{i\omega t}) \,.
\end{equation}

The Schrödinger equation for the resulting state $|\psi_0(t)\rangle$ reads:
\begin{equation}
    i\hbar \frac{d}{dt} \psi_0(x,t)= \left[\frac{\hat p^2}{2m} + V(\hat x-\hat \alpha(t)) \right] \psi_0(x,t).
\end{equation}

As in the previous Section,  we shift the position $\hat x$ to $\hat \alpha(t)$, using the unitary transformation
\begin{equation}
       \hat U_0 = \exp\left(\frac{\partial}{\partial x} \mathcal{T}\int_{t_0}^t dt'\,\dot{\hat\alpha}(t') \right),
\end{equation}

As standard in quantum field theory (QFT) or scattering theory, we explicitly suppress the coupling at initial time.
This time
\begin{equation}
    \dot{\hat \alpha}(t) = i\int d\omega\, \ell_\omega \, \omega \left(-\hat a_\omega e^{-i\omega t}+\hat a_\omega^\dagger e^{i\omega t}\right)\,.
\end{equation}
As before, we obtain,
$
    \hat U_0 \hat x \hat U_0^\dag = \hat x + \hat\alpha(t).
$

The resulting Schrödinger equation for $|\psi_1(t)\rangle=\hat U_0 |\psi_0(t)\rangle$ reads:
\begin{equation}
    \begin{aligned}
    &i\hbar \frac{d}{dt} \psi_1(x,t)
    = \left[\frac{(\hat p-m \dot {\hat\alpha}(t) )^2}{2m} -\frac{m \dot {\hat \alpha}(t)^2}{2}+ V(\hat x)\right]\psi_1(x,t).
    \end{aligned}
\end{equation}

We shall follow exactly the same steps as in Sec.~\ref{secIII}. We introduce $|\psi_2(t)\rangle = \hat U_{1}|\psi_1(t)\rangle$, where
\begin{equation}
    \hat U_{1}=\mathcal{T}\exp\left(-\frac{im}{2\hbar}\int_{t_0}^t dt'\,  \dot{\hat \alpha}(t')^2\right) ,
\end{equation}
getting
\begin{equation}
    \begin{aligned}
    &i\hbar \frac{d}{dt} \psi_2(x,t) 
    = \left[\frac{(\hat p-m \dot {\hat\alpha}_\text{eff}(t) )^2}{2m} + V(\hat x)\right]\psi_2(x,t).
    \end{aligned}
\end{equation}

Finally, we use $|\psi(t)\rangle=\hat U_2 |\psi_2(t)\rangle$, with
\begin{equation}
    \hat U_2=\exp\left(\frac{-im }{\hbar} \hat x \, \mathcal{T}\int_{t_0}^t dt'\,\ddot{\hat \alpha}_\text{eff}(t') \right) \,,
\end{equation}
which brings us to the final result
\begin{equation}
i\hbar \frac{d}{dt} \psi(x,t) = \left[\frac{\hat p^2}{2m} + V(\hat x) + m \hat x \ddot{\hat\alpha}_\text{eff}(t)\right]\psi(x,t).
\label{final2}
\end{equation}

The remaining task is to calculate $\hat\alpha_\text{eff}(t)$, which is realized in full analogy to Sec.~\ref{secIII},
\begin{equation}
    \begin{aligned}
    &\exp\!\left(\frac{-i m \dot{\hat \alpha}(t')^2}{2\hbar}\right)\ \dot{\hat \alpha}(t)\exp\!\left(\frac{i m \dot{\hat \alpha}(t')^2}{2\hbar}\right)
=  \dot{\hat \alpha}(t) - \frac{i m}{2\hbar}\left[\dot{\hat \alpha}(t')^2, \dot{\hat \alpha}(t)\right]\\ 
    \\
    &\hspace{1cm}=  \dot{\hat \alpha}(t) - \frac{i m}{\hbar}F(t',t)\dot{\hat \alpha}(t'). 
    \end{aligned}
\end{equation}

In the continuous spectrum case, we have
$F(t',t)= 2 i \int d\omega\, \ell_\omega^2\omega^2\sin[\omega(t-t')]$. Still, the final expressions are analogous
\begin{equation}
    \begin{aligned}
     \dot{\hat\alpha}_\text{eff}(t)&= \hat U_{1} \dot{\hat \alpha}(t) \hat U_{1}^\dag
     = \dot{\hat\alpha}(t) - \frac{im}{\hbar}\int_{t_0}^t dt' \, F(t', t)\dot{\hat \alpha}(t')+ \rm{...}
    =\dot{\hat\alpha}(t)\left[1 + O(\epsilon) + O(\epsilon^2) + ...\right].
    \label{alphaR2}
    \end{aligned}
\end{equation}

In parallel to Section \ref{secIII}, the main result of the present Section is contained in Eqs.~\eqref{final2} and \eqref{alphaR2}. As before, the quantum KH transformation leads to the Schrödinger equation for the trapped particle in the quantized electric field, with operator ${\cal E}(t) = -\frac{m}{e}\ddot{\hat \alpha}_\text{eff}(t)$. Interestingly, this effective field is not simply equal to the mean value $\expval*{\ddot{\hat \alpha}(t)}$, as one might expect from the classical analysis; rather, it includes a quantum electrodynamic correction, as illustrated in Eq. (\ref{alphaR2}). We now numerically benchmark the consequences of these results.

\section{Numerical evaluation}
\label{secNum}

\begin{figure*}[!tbp]
    \centering
    \includegraphics[width=0.75\linewidth]{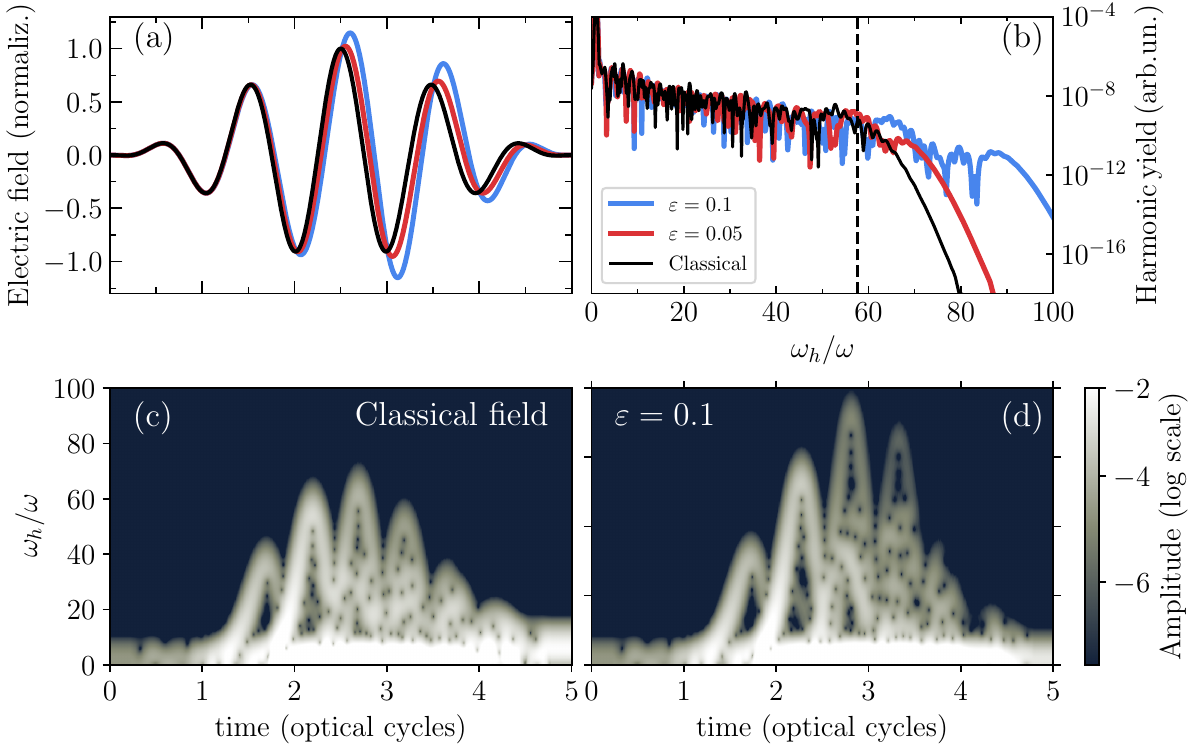}
    \caption{
    (a) Average value of the electric field at different times of the optical cycle, for the classical input (black), and a first-order of quantum correction (color, see main text). (b) HHG yield as a function of different multiples of the fundamental frequency $\omega=0.057$ a.u. (corresponding to a laser wavelength $\lambda=800$ nm). Gabor transform of the harmonic yield for the classical input (c) and a correction strength, $\epsilon=0.1$ (d).}
    \label{fig:numericalHHG}
\end{figure*}

In Sec.~\ref{secIII}, we observed that the quantization of the position of the moving trap results in corrections of the effectively applied field.~The parameter $\epsilon$ defined in Eq.~\eqref{eq:def_epsilon} dictates the strength of these corrections. As a particular example, here we evaluate their impact in the application of a sinusoidal $n_c$-cycle pulse, 
\begin{equation}
    f(t)=\sin^2\left(\frac{\omega t}{2 n_c} \right)
\end{equation}
commonly used in the context of HHG in atoms and solids.

As an initial exploration of the corrections introduced by the quantized shaking of the trap, we consider the case of an oscillatory coherent state. Taking average values $\ev{\hat a}=|\upalpha| e^{i\phi}$ in Eq.~\eqref{eq:corrections_1freq}, and expanding up to corrections of order $O(\epsilon)$, we obtain an effective field,
\begin{equation}
\label{eq:aeff_first_order}
    \begin{aligned}
    &\ev{\ddot{\hat\alpha}_\text{eff}(t)} \simeq
        -2\abs{\upalpha}\ell \omega^2  f(t)
        \bigg[
            \cos(\omega t-\phi)
+ 2 \epsilon \,\omega\!
        \int^{t}_{t_0}\!\!\dd t'
            f^2(t')\cos[\omega (t-t')]\sin(\omega t' - \phi)
        \bigg].
    \end{aligned}
\end{equation}
Therefore, the quantum movement of trap introduces a correction in the coherent amplitude of the field experienced by the trapped particle that results from the non-commutativity of the position operator at previous times. In Figure~\ref{fig:numericalHHG}(a) we compare corrections of order $\epsilon \sim 0.1$ (coloured lines) to the case of a classically moving trap (black). We observe an increased intensity of the field that is more noticeable around the last cycles of the pulse, where the position of the maxima gets shifted and there is an effective down-chirp resulting from the accumulative effect of the correction. Since tunnel ionization is extremely sensitive to the instantaneous field strength, this late-time increase can strongly change which release times contribute most to HHG and can shift the return times of the usual short and long trajectories.

These changes in the waveform also modify the phase accumulated by the electron in the continuum. As a result, the relative phase between different trajectories (and between successive half-cycles) is altered, so the harmonic emission can become more or less constructive at specific orders. In this way, a modest 
$O(\epsilon)$ correction can noticeably reshape the HHG spectrum by changing both trajectory weights and their interference.

Likewise, in Figure~\ref{fig:numericalHHG}(b), we plot the HHG spectra for a target with $I_p=0.5$ a.u. (simulated Hydrogen), illuminated with an $n_c=5$ pulse of frequency $\omega=0.057$ a.u. and intensity $I=4\times10^{14}$ W/cm$^{2}$, corresponding to a electric field peak strength $E_0=0.107$ a.u (see numerical details in Appendix~\ref{numerics}). The temporal profile of the driving electric field follows the time-dependent waveforms shown in Figure~\ref{fig:numericalHHG}(a), i.e., we include the corrections of order $O(\epsilon)$ present in Eq.~\eqref{eq:aeff_first_order} to the bare electric field. Interestingly, we observe a substantial extension of the HHG cutoff frequency, $\omega_c/\omega\sim 57$, that the three-step model predicts for a classical input field~\cite{LBI1994}. 

Such an extension is expected from the enhanced amplitude of the corrected field, which increases the ponderomotive energy of the emitted particle. In addition to this enhancement, the modified waveform does more than simply rescale the overall yield: it reshapes the sub-cycle return map that links ionization time to recollision energy. In the three-step picture, the cutoff is set by the largest kinetic energy available at return but, in practice, that maximum is accessed only during a narrow window of release phases. A small enhancement of the field near the relevant half-cycles can shift and broaden these optimal windows, increasing the weight of high-energy returns and effectively pushing the observable cutoff to higher orders.

Equally important, the cutoff region is governed by interference between a small number of dominant pathways. The harmonic signal at a given frequency is a coherent sum over contributions with different excursion times, so changing the field modifies the accumulated action and the relative phase between short and long trajectories, as well as between successive half-cycle bursts. Near the classical cutoff, where phases vary rapidly and contributions are already marginal, even modest phase shifts can turn near-cancellation into partial constructive interference, making higher-order components visible. In this way, the observed cutoff extension can be understood as a combined effect of altered trajectory energetics and altered trajectory interference, both controlled by the detailed temporal structure of the corrected driving field.

This interpretation is further supported by the time–frequency analysis.~The Gabor transforms~\cite{gabor1946theory} of the HHG spectra shown in Fig.~\ref{fig:numericalHHG}(c,d) illustrate that frequencies beyond the cutoff mostly originate from the maxima present after $t\simeq2.5$ optical cycles, which are favored by the `quantum' corrections.~In the trajectory language, this means that the `extra' high-order emission is not produced uniformly throughout the pulse, but is gated to specific half-cycle regions where the `QKH' corrections act more strongly, and therefore where the instantaneous field and the associated vector potential combine to launch and drive electrons into the most energetic recollisions. 

This kind of temporal localization is exactly what one expects when small waveform changes selectively amplify a family of quantum orbits. The relevant trajectories are those whose recombination times lie near the field maximum around $t\simeq3$ cycles, with the return occurring after an excursion time that maximizes the recombination energy; strengthening the family of quantum orbits compatible with such recombination times increases their tunneling amplitude and can also shift their recollision conditions to slightly higher energies. At the same time, concentrating the high-frequency emission into a narrower temporal window reduces competition with other half-cycle bursts, so destructive interference from neighboring contributions is weakened in the cutoff region. For instance, for harmonic orders $\gtrsim 80$ only a single trajectory appears to contribute, making this spectral region uniform. The Gabor map therefore connects the spectral observation (new components beyond the classical cutoff) to a concrete dynamical mechanism: the quantization-induced corrections act as a sub-cycle gate that favors the specific burst responsible for the most energetic—and hence highest-frequency—recollisions.

\section{Experimental proposal}
\label{secVI}
\begin{figure}[tb]
    \centering
    \includegraphics[width = 0.4\columnwidth]{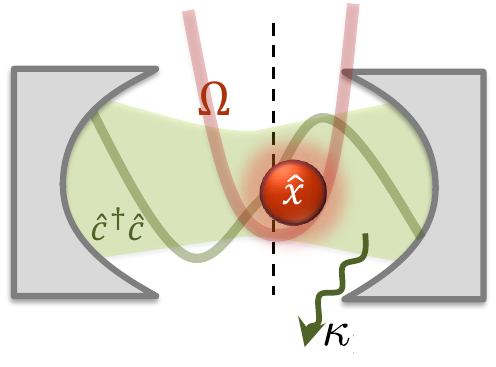}
\caption{Optomechanical scheme where the position $\hat x$ of an atomic cloud (red) inside a gaussian optical trap with mechanical frequency $\Omega$ is affected by the optomechanical interaction in Eq.~\eqref{eq:hom} with a cavity mode (green) populated with $\langle \hat c^\dagger \hat c \rangle$ average photons, and a decay rate $\kappa$.}
\label{fig:scheme}
\end{figure}

One possible way to mechanically control the quantum position of the trap is to do it optically. This is the subject of optomechanics, where the mechanical position $\hat x$ of a resonator is controlled by a photonic field that has associated annihilation (creation) operators, $\hat c^{(\dagger)}$, and an optical frequency shift per displacement $G$ of the form~\cite{Aspelmeyer2014}:
\begin{equation}
\label{eq:hom}
    \hat H\stx{OM}=\hbar G \,\hat x \hat c^\dagger \hat c \,. 
\end{equation}
This resonator of mass $m$ can correspond, for example, to a suspended mirror shined by a laser~\cite{Aspelmeyer2014,kippenberg2008cavity}, or an atomic system trapped in a cavity~\cite{ritsch2013cold,stamper2014cavity,murchObservation2008,arguello2022optomechanical}, as represented in Fig.~\ref{fig:scheme}.~Expanding the trapping potential around its center as a quadratic potential with frequency $\Omega$, $V(\hat x)=m\Omega^2\hat x^2/2$, the total optomechanics Hamiltonian reads,
\begin{equation}
\label{eq:OM}
\begin{split}
    \hat H =\frac{\hat p^2}{2m} + \frac{m\Omega^2}{2} \pa{\hat x + \frac{\hbar G}{m\Omega^2}\hat c^\dagger \hat c }^2-\frac{\hbar^2 G^2}{2m\Omega^2}(\hat c^\dagger \hat c)^2\,.
    \end{split}
\end{equation}
In the case of coherent or displaced squeezed states~\cite{gardiner2004quantum}, it is relevant to define the quantum fluctuations of the photonic field around its average value, $\hat a\equiv \hat c-\ev{\hat c }$, where we consider $\ev{\hat c}=\sqrt{n_{ph}}$ to be real valued and related to the average number of photons in the cavity. 

The latter term in Eq.~\eqref{eq:OM} can be eliminated with a unitary transformation $\hat U_G=\exp\co{-i\hbar G^2 (\hat c^\dagger \hat c)^2 t/( 2m\Omega^2)}$, so that one is left with an optomechanical Hamiltonian that identifies with the description of Eq.~\eqref{eq:quantizedShaking}, for a quadratic potential $V(\hat x)$ and a quantized center of the trap $\hat \alpha_x(t)=\alpha_m(t)+\hat\alpha_m(t)$, which has a classical position defined by the cavity population $\alpha_m(t)=\ell f(t)\sqrt{n\stx{ph}(t)}$, and a quantized term dictated by the photonic operator, $\hat\alpha_m(t)=\ell f(t) (\hat a+\hat a^\dagger)$. Following this mapping, one identifies
\begin{equation}
    \ell f(t)=\hbar G \sqrt{n\stx{ph}(t)} /(m\Omega^2)=2x_{zp}\sqrt{n\stx{ph}(t)} g_0/\Omega\,,
\end{equation}
where $g_0=G x\stx{zp}$ is the optomechanical single-photon coupling strength, and $x\stx{zp}=\sqrt{\hbar/(2m\Omega)}$ is the zero-point motion of the resonator.

The parameter that accounts for nonlinear terms emerging in the QKH transformation now reads 
\begin{equation}
    \epsilon=\frac{\ell^2}{2a_\text{zp}^2}=2n\stx{ph}\frac{g_0^2}{\Omega^2}\frac{\omega}{\Omega}\,.
\end{equation}

One possible optomechanical implementation is an ultracold atomic cloud~\cite{brenneckeCavity2008,murchObservation2008,schleier-smithOptomechanical2011} trapped in a cavity~\cite{Aspelmeyer2014}. The center of mass of the cloud plays the role of the mechanical degree of freedom, which is optically controlled by the coupling to a spatially-dependent cavity mode. One can consider typical values $\Omega=2\pi\times 400$ kHz, an effective coupling $g_0\sim 2\pi\times 100$ Hz for $^{87}$Rb atoms. The relevant driving frequency is controlled by the detuning with the cavity frequency, which can be chosen as $\omega\sim 100$ MHz. For a photonic population, $n\stx{ph}\sim 10^{4}$, the associated expansion parameter is $\epsilon\sim 0.1$, and the fluctuations of the cavity field should have an effect on the atomic dynamics.~In the context of AS, one can choose an input field that modulates the average number of photons as $n\stx{ph}(t)=n_0\co{1-\sin(\omega_0 t)}$. Inducing such modulation of the effective force requires that the induced variations are slower than those dictated by the cavity decay parameter, $\omega_0 \ll \kappa \sim 2\pi\times 1$ MHz, so that the photonic population gets adjusted accordingly. 

This experimental implementation would require accounting for imperfections that should be analyzed in detail for each particular simulation in order to optimize their performance. Some of the most relevant ones are: (i) Cavity dissipation injects quantum fluctuations into the system that impart a random radiation pressure force on the atoms. Non-adiabatic retardation effects also need to be taken into consideration when engineering the effective field inside the cavity. (ii) The atoms are subjected to scattering events that introduce recoil heating and decoherence. Finite temperature would also populate higher motional states of the trap, which correspond to excited states of the simulated target. (iii) The laser intensity and phase noise translate into classical fluctuations of the photon number and the detuning with the cavity resonance, respectively. Overall, these are challenges that are already faced by the cavity QED community and can benefit from their ongoing efforts to characterize and mitigate their impact in the simulation.

\section{Consequences and outlook}
\label{secV}

The main goal of this work was to derive the quantum KH transformation, which we have demonstrated for the illustrative single-mode case (Section \ref{secIII}), and in the case of a continuous spectrum (Section \ref{secIV}). From this point, one relevant question ahead is: what kind of effects and phenomena can one observe with these kinds of quantum simulators of QED and QO of AS? 

While in standard experiments of AS one can measure the photons in the fundamental laser field or the generated harmonics, such measurements require a careful design when working with quantum simulators of QED and QO of AS each of. Such analysis goes far beyond the scope of the present paper; nevertheless, in the following, we present a list of possible directions for future research. 
 For concreteness, we assume here that the initial state of our systems is the ground state of the particle in the quantum mechanical trap, and a Gaussian state of the quantum oscillator, e.g. a coherent or squeezed state.

\begin{itemize}
\item Interestingly, even if one starts with with a classical coherent state and performs a semi-classical analysis (replacing the associated field operator by its quantum mechanical average), quantum effects emerge from the QED correction to the effective field [Eqs.~\eqref{alphaR} and \eqref{alphaR2}]. This effect could be measured by monitoring `ionization rates', which are independent of the observation frame. This is, the escape rate from the trap, as it was proposed in the case of quantum simulators using classical KH transformation in Refs.~\cite{DSS98,AH10, SFS17, RFS17,SRS18}.
\item In addition, one can also look at the effects that QED corrections impose in the effective `electric' field, using the methods proposed for quantum simulation of HHG in~\cite{ARS24} or EII~\cite{Maribel} in optical-tweezer traps, which can be shaken or controlled by external time-dependent EM fields. 
\item When one aims to generate MQS using a quantum simulator, the situation becomes more demanding. Similarly to the conventional scenario where depletion (escape from the trap) is negligible, one can use post-selection methods for this goal, which requires certain control of the quantum mechanical state of the quantum oscillator. Such approach is compatible with current technology and physical ideas~\cite{LCP21,RSP21,SRL22,rivera-dean_bloch_2023,Stammer_ArXivConditioning_2024,Stammer_EnergyConservation2024}, but requires concrete designs aimed form specific physical systems.
\item Finally, when depletion is non-negligible, one expects that the state of the quantum simulator will naturally evolve to a MQS, typically a multi-mode squeezed state.~Here, the physics will be analogous to the generation of entangled states including dipole correlations beyond the low depletion limit as shown in Ref.~\cite{Stammer_PRL_2024}, and to the generation of light-matter entanglement in atoms, molecules, and semiconductors~\cite{RLP22,RSM22,Riv24,rivera-dean_bloch_2023,CKN24}.
\end{itemize}

Going beyond these ideas, there are two more interesting generalizations that can open new avenues for quantum simulation.~The first is to go beyond one-dimensional systems and also try to simulate quantum structured light, including elliptically polarized quantum light, or more interesting structured lights that combine driving fields at different frequencies~\cite{PJV19,PRS19,RDB19}.
The second consists of the control of the quantum state of the mechanical oscillator, which would allow in principle to prepare it in a quantum state, such as a squeezed state, or even more exotic ones. Combining squeezing with structured light might lead to classical forbidden results, as recently shown in Ref.~\cite{RSC24}.

\begin{acknowledgments}
ICFO-QOT group acknowledges support from: 
MCIN/AEI (PGC2018-0910.13039/501100011033,  CEX2019-000910-S/10.13039/501100011033, Plan National STAMEENA PID2022-139099NB, project funded MCIN and  by the “European Union NextGenerationEU/PRTR" (PRTR-C17.I1), FPI); Ministry for Digital Transformation and of Civil Service of the Spanish Government through the QUANTUM ENIA project call - Quantum Spain project, and by the European Union through the Recovery, Transformation and Resilience Plan - NextGenerationEU within the framework of the Digital Spain 2026 Agenda; CEX2024-001490-S [MICIU/AEI/10.13039/501100011033]; Fundació Cellex;
Fundació Mir-Puig; Generalitat de Catalunya (European Social Fund FEDER and CERCA program; Barcelona Supercomputing Center MareNostrum (FI-2023-3-0024); 
Funded by the European Union (HORIZON-CL4-2022-QUANTUM-02-SGA, PASQuanS2.1, 101113690, EU Horizon 2020 FET-OPEN OPTOlogic, Grant No 899794, QU-ATTO, 101168628),  EU Horizon Europe Program (No 101080086 NeQSTGrant Agreement 101080086 — NeQST).
J.A.-L. acknowledges support from the Spanish Ministerio de Ciencia, Innovación y Universidades (grant PID2023-147469NB-C21, financed by MICIU/AEI/10.13039/501100011033 and FEDER-EU).
P.S. acknowledges funding from the European Union’s Horizon 2020 research and innovation program under the Marie Skłodowska-Curie grant agreement No 847517.
M.~F.~C.~acknowledges support by the Quantum Science and Technology-National Science and Technology Major Project (Grant No. 2025ZD0301000),  the National Key Research and Development Program of China (Grant No. 2023YFA1407100), the Guangdong Province Science and Technology Major Project (Future functional materials under extreme conditions - 2021B0301030005) and the National Natural Science Foundation of China (Grant No. 12574092).
\end{acknowledgments}

\providecommand{\noopsort}[1]{} \providecommand{\noopsort}[1]{}

\newpage
\clearpage
\appendix
\begin{center}
    {\large {\bfseries\scshape Appendix}}
\end{center}
\section{Explicit evaluation of $\dot{\hat{\alpha}}_\text{eff}(t)$}
\label{sec:appendix}

In this Appendix we provide a more detailed derivation of $\dot{\hat{\alpha}}_\text{eff}(t)$, including explicit expressions for the higher-order terms. Formally, this operator is defined as
\begin{equation}\label{Eq:App:alpha:R}
	\dot{\hat{\alpha}}_\text{eff}(t)
		= \hat{U}_{1}(t,t_0)
				\dot{\hat{\alpha}}(t)
			\hat{U}_{1}^\dagger (t,t_0), 
\end{equation}
for the time-evolution operator $
	\hat{U}_{1}(t,t_0)
		= \mathcal{T}
			\exp[
				-\tfrac{im}{2\hbar} \int^{t}_{t_0}
					 \dd \tau \dot{\hat{\alpha}}(\tau)^2]
$.~We note that this time-evolution operator can also be factorized as
\begin{equation}\label{Eq:App:Partition}
	\begin{aligned}
	\hat{U}_1(t,t_0)
		&= \prod_{j=1}^{N}
			\exp[-\dfrac{im}{2\hbar}  \dot{\hat{\alpha}}(t_j)^2 \Delta t]
		\\&= \hat{U}_1(t_N)\hat{U}_1(t_{N-1}) \cdots
			\hat{U}_1(t_2)\hat{U}_1(t_{1}),
	\end{aligned}
\end{equation}
by defining $\hat U(t_j)=\hat U(t_j,t_{j-1})$ for time intervals $\Delta t = (t-t_0)/N$, with $N$ sufficiently large.
Using Eq.~\eqref{Eq:App:Partition}, we can rewrite Eq.~\eqref{Eq:App:alpha:R} as
\begin{equation}
	\dot{\hat{\alpha}}_\text{eff}(t)
	= \hat{U}_1(t_N) \cdots
			\hat{U}_1(t_{1})
				\dot{\hat{\alpha}}(t)
			\hat{U}_1^\dagger(t_1)
		\cdots \hat{U}_1^\dagger(t_{N}),
\end{equation}
where, in what follows, we define $\dot{\hat{\alpha}}^{(j)}_\text{eff} =  \hat{U}_1(t_j) \cdots  \hat{U}_1(t_1)\dot{\hat{\alpha}}(t)\hat{U}_1^\dagger(t_1)\cdots \hat{U}_1^\dagger(t_{j})$.~To gain insight into the structure of this expression, we explicitly write it out for a few iterations. For this purpose, it is particularly convenient to use the Baker-Campbell-Hausdorff (BCH) formula~\cite{GerryBookCh1}, which for two general operators $\hat{X}$ and $\hat{Z}$ reads
\begin{equation}
	e^{-i \hat{Z}}
		\hat{X}
	e^{i \hat{Z}}
		= \hat{X}
			- i[\hat{Z},\hat{X}]
			- \dfrac{1}{2!}
				\big[\hat{Z},[\hat{Z},\hat{X}]\big]
			+ \cdots	
\end{equation}

In our case, we identify $\hat{Z} \equiv \beta\dot{\hat{\alpha}}(t_j)^2/2$, and $\hat{X} \equiv \dot{\hat{\alpha}}(t)$, with $\beta=m\Delta t/\hbar$. Thus, we find for the relevant commutators in the BCH expansion
\begin{align}
	&[\hat{Z},\hat{X}]
		= \beta F(t_j,t)  \dot{\hat{\alpha}}(t_j),
	\\& \big[\hat{Z},[\hat{Z},\hat{X}]\big] = 0,
\end{align}
with the higher-order terms vanishing as well. Here, $F(t',t) = \left[\dot{\hat \alpha}(t'), \dot{\hat \alpha}(t)\right]$ is a $c$-number. Consequently, we obtain
\begin{align}
	\dot{\hat{\alpha}}^{(1)}_\text{eff}(t)
		&= \dot{\hat{\alpha}}(t)
			- i\beta F(t_1,t) \dot{\hat{\alpha}}(t_1)\,,\\
	\dot{\hat{\alpha}}^{(2)}_\text{eff}(t)
		&= \dot{\hat{\alpha}}(t)
			- i\beta F(t_1,t) \dot{\hat{\alpha}}(t_1)
			- i\beta F(t_2,t) \dot{\hat{\alpha}}(t_2) \nonumber \\
		& 	- \beta^2 F(t_2,t_1)F(t_1,t)\dot{\hat{\alpha}}(t_2).
\end{align}

From these two examples, we can already gain insight into the general transformation: higher order iterations of $j$ not only reproduce the same orders of $i\beta$ as those appearing in the $(j-1)$th iteration, but also introduce additional contributions that are multiplied by chained products of commutators $F(t_1,t)\prod^{j-1}_{k=1}F(t_{k+1},t_{k})$.~Consequently, in the limit $\delta \to 0$ (equivalently, $N \to \infty$) they can be expressed as a series of time-ordered integrals:
\begin{equation}
	\begin{aligned}
	\dot{\hat{\alpha}}_\text{eff}(t)
		&= \dot{\hat{\alpha}}(t)
			-\dfrac{im}{\hbar}
				\int^{t}_{t_0} 
					\dd \tau_1 F(\tau_1,t)
						\dot{\hat{\alpha}}(\tau_1)
			\\&\quad
			+	\pa{\dfrac{im}{\hbar}}^2
				\int^{t}_{t_0} \!\!\dd \tau_2
					\int^{\tau_2}_{t_0} \!\!\dd \tau_1
						F(\tau_2,\tau_1)F(\tau_1,t)
								\dot{\hat{\alpha}}(\tau_2)
			\\&\quad+ \cdots
	\end{aligned}
\end{equation}
Thus, in the main text we identify 
\begin{align}
	 O(\epsilon)
		&= -\dfrac{im}{\hbar}
			\int^{t}_{t_0} 
				\dd \tau_1 F(\tau_1,t)
					\dot{\hat{\alpha}}(\tau_1),
	\\ O(\epsilon^2)
		& = \bigg(
				\dfrac{im}{\hbar}
			\bigg)^{\!2}\!\!
			\int^{t}_{t_0} \!\!\dd \tau_2
				\int^{\tau_2}_{t_0} \!\!\dd \tau_1
						F(\tau_2,\tau_1)F(\tau_1,t)
							\dot{\hat{\alpha}}(\tau_2).
\end{align}

\section{Numerical evaluation of QKH corrections in HHG}
\label{numerics}
The numerical scheme is based on direct real-time propagation of the one-dimensional time-dependent Schrödinger equation (1D-TDSE) on a uniform spatial grid using an implicit Crank–Nicolson integrator, which preserves unitarity and stability under strong driving time-dependent sources. The atomic Hamiltonian is discretized with second-order finite differences, yielding a tridiagonal structure that is solved efficiently at each time step via banded linear algebra. The laser–matter interaction is treated in the length gauge, optionally including a QKH-like correction term, and absorbing boundary layers are applied to suppress reflections at the spatial grid boundaries. The ground state is obtained from the lowest eigenpair of the field-free tridiagonal Hamiltonian and used as the initial condition. During propagation, the high-harmonic signal is computed from the dipole acceleration $a(t)=-\langle \psi(t)|\nabla V|\psi(t)\rangle$, sampled at uniform intervals to ensure FFT compatibility. The HHG spectrum is then obtained through Fourier transformation of $a(t)$, while the time–frequency dynamics are resolved using a Gabor (Gaussian windowed Fourier) transform, providing a joint temporal–spectral representation that reveals emission bursts, cutoff formation, and trajectory interference within a consistent TDSE-level framework.
\end{document}